\documentclass[a4paper]{article}
\RequirePackage[english]{babel}
\RequirePackage[latin1]{inputenc}
\RequirePackage[T1]{fontenc}
\RequirePackage{mathrsfs}
\RequirePackage{amsmath}
\RequirePackage{amssymb}
\RequirePackage{amsbsy}
\RequirePackage{bm}
\begin{document}
\title{\bf{From the Torsion Tensor for Spinors\\ to the Weak Forces for Leptons}}
\author{Luca Fabbri\\ \footnotesize  Dipartimento di Fisica, Universit\`a di Bologna, 
Via Irnerio 46, 40126 Bologna, ITALY}
\date{}
\maketitle
\begin{abstract}
We consider a geometric approach to field theory in which torsion is present beside gravity and also electrodynamics for the matter field equations, and we develop the consequences of the torsion-spin coupling for the spinor fields; we show that these interactions have the structure of the weak interactions acting among leptons: we discuss the implications for the standard model of fundamental interactions of elementary fields in the perspective of the foundations of unification in theoretical physics.
\end{abstract}
\section*{Introduction}
In the Weinberg-Salam standard model \cite{f/0}, the weak forces are the result of their separation from electrodynamics after the electro-weak interactions have been disunified, while the weak massive mediators have gotten their masses via the generation of the mass of initially massless vector bosons; the disunification of the initially unified electro-weak interactions is given by breaking the symmetry of the underlying $SU(2)\times U(1)$ group, while the generation of the masses of the initially massless vector bosons is given by transferring additional degrees of freedom into the gauge potentials of the $SU(2)\times U(1)$ group itself; the transfer of degrees of freedom is operated by exploiting gauge transformations, and before this transfer the degrees of freedom are stored in the Higgs scalar: the breaking of gauge symmetry, generating the masses of the gauge fields, is due to the presence of a potential for the Higgs field itself. This model has three types of problems: from a theoretical perspective, the $SU(2)\times U(1)$ group is not really a unification but only a patching of the two $SU(2)$ and $U(1)$ sub-groups; from a phenomenological perspective, the generation of the masses of the leptons is given in terms of the Yukawa potential so that the origin of the masses of leptons is no more explained than the origin of the Yukawa couplings; from an experimental perspective, the Higgs field has not been detected yet. The most straightforward way in which all these problems may be avoided would be to look for models in which there is no Higgs field, and therefore no mass generation, nor breaking of any symmetry possible; in this Higgsless configuration there would be unified massless electro-weak interactions. This model would still have two kinds of problems; phenomenologically, the generation of the masses of leptons would now be unknown, experimentally, these electro-weak interactions have not been observed. Again the most direct way in which both these problems may be circumvented would be to look for models in which there is no Higgs field nor even electro-weak interactions. And again this model would still have a problem, as in experiments massless leptons are unseen. The way to solve this problem is to postulate fundamental fermions to be already massive, interacting only with electrodynamics and eventually gravity, to ensure the most general coupling possible. But now the question is, given that the weak forces could not be fundamental, symmetric and massless fields, what can recover their effects?

Now, in the most general dynamics for matter fields, the most general derivative for fermion fields is given with respect to the most general spinorial connection: the most general spinorial connection can be decomposed in terms of an abelian gauge field plus the most general tensorial connection: in this way the most general tensorial connection can be decomposed into the metric connection and torsional degrees of freedom; thus the most general connection already accounts for electrodynamics, gravitation and torsion, as it has been discussed in reference \cite{s-s}. In this connection, both the fields of gauge and the metric are such that their vanishing depends on the gauge and frame we employ, but torsion never vanishes: therefore torsion can always be separated away, as discussed in reference \cite{h-h-k-n}; after this decomposition, the field equations in the torsional case reduce to the field equations in the torsionless case plus additional torsional contributions that can be written as additional interactions for the spinors, as in references \cite{s-s/1} and \cite{h-d}. In the Dirac equation these additional interactions between spinors are actually interactions between spinorial right-handed and left-handed projections, thus providing effective mass terms and self-interactions for each spinor and mutual interactions for pairs of spinors with each other, as in \cite{s-g} and \cite{s-s/2} and also \cite{f/1}; at this point we may therefore ask, given these torsional interactions, what could their effects be?

These two questions may have a hand-glove answer, that is torsional interactions may be what induces the effects of the weak forces. In the present paper, we shall start from the most general Dirac field equations, we will isolate torsional contributions and we are going to show that the torsional interactions for spinors are formally identical to the weak interactions for leptons, given by massive and fixed, composite degrees of freedom.
\section{From Torsion Tensor for Spinors\\ to Weak Forces for Leptons}
In the present paper we shall refer to \cite{f/1} for the notations and the general formalism and also the definition of fundamental quantities, and accordingly we will consider the most general matter field equations to be given by
\begin{eqnarray}
&i\gamma^{\mu}D_{\mu}e=me\\
&i\gamma^{\mu}D_{\mu}\nu=0
\end{eqnarray}
where the matrices $\gamma_{\mu}$ belong to the Clifford algebra and in which $D_{\mu}$ is the most general covariant derivative with respect to the most general connection containing torsion and metric and also gauge potentials, whose action is defined on both the electron $e$ having charge $q$ and mass $m$ and the neutrino $\nu$ being neutral and massless and consequently chosen to be in this case single-handed.

In order for these matter field equations to be simplified, we will also introduce the parity-odd matrix $\gamma=i\gamma^{0}\gamma^{1}\gamma^{2}\gamma^{3}$ and we further remark that the most general covariant derivative $D_{\mu}$ can be decomposed by separating the most general connection into the simplest connection plus the torsion tensor thus giving the simplest covariant derivative $\nabla_{\mu}$ plus the torsional contributions as
\begin{eqnarray}
&i\gamma^{\mu}\nabla_{\mu}e
-\frac{3}{16}\overline{e}\gamma_{\mu}e\gamma^{\mu}e
-\frac{3}{16}\overline{\nu}\gamma_{\mu}\nu\gamma^{\mu}\gamma e=me\\
&i\gamma^{\mu}\nabla_{\mu}\nu
-\frac{3}{16}\overline{e}\gamma_{\mu}\gamma e\gamma^{\mu}\nu=0
\end{eqnarray}
having used for the double-handed spinor $\overline{e}\gamma_{\mu}\gamma e\gamma^{\mu}\gamma e 
+\overline{e}\gamma_{\mu}e\gamma^{\mu}e\equiv0$ and for the single-handed spinor $\overline{\nu}\gamma_{\mu}\nu\gamma^{\mu}\nu=\overline{\nu}\gamma_{\mu}\gamma\nu\gamma^{\mu}\gamma\nu\equiv0$ and in which the contributions of torsion are interactions between right-handed and left-handed projections of all spinors, resulting in effective mass terms and self-interactions for the spinor plus mutual interactions for the couple of spinors with one another; the task we now have to reach is therefore to write these torsional interactions among spinors in the form of the weak interactions among leptons.

To further pursue the treatment of the field equations so decomposed, we are now going to decompose these spinors in their right-handed and left-handed projections $2e_{R}=(\mathbb{I}+\gamma)e$ and $2e_{L}=(\mathbb{I}-\gamma)e$ with $\nu_{R}=0$ and $\nu_{L}=\nu$ and then we are going to employ the Fierz identities to see that the above equations can be rearranged in the equivalent form
\begin{eqnarray}
\nonumber
&i\gamma^{\mu}\nabla_{\mu}e
-\frac{3}{8}(\cos{\theta})^{2}\overline{e}\gamma_{\mu}e\gamma^{\mu}e+\\
\nonumber
&+q\tan{\theta}\left[\frac{3\cos{\theta}}{8g\left(\sin{\theta}\right)^{2}}
\left[\frac{1}{2}\left(\overline{e}_{L}\gamma_{\mu}e_{L}-\overline{\nu}\gamma_{\mu}\nu\right)
-(\sin{\theta})^{2}\overline{e}\gamma_{\mu}e\right]\right]\gamma^{\mu}e-\\
\nonumber
&-\frac{g}{2\cos{\theta}}\left[\frac{3\cos{\theta}}{8g\left(\sin{\theta}\right)^{2}}
\left[\frac{1}{2}\left(\overline{e}_{L}\gamma_{\mu}e_{L}-\overline{\nu}\gamma_{\mu}\nu\right)
-(\sin{\theta})^{2}\overline{e}\gamma_{\mu}e\right]\right]\gamma^{\mu}e_{L}+\\
&+\frac{g}{\sqrt{2}}
\left[\frac{3\sqrt{2}\left[1-4(\sin{\theta})^{2}\right]}{32g\left(\sin{\theta}\right)^{2}}
\overline{e}_{L}\gamma_{\mu}\nu\right]^{*}\gamma^{\mu}\nu=me\\
\nonumber
&i\gamma^{\mu}\nabla_{\mu}\nu
+\frac{g}{2\cos{\theta}}\left[\frac{3\cos{\theta}}{8g\left(\sin{\theta}\right)^{2}}
\left[\frac{1}{2}\left(\overline{e}_{L}\gamma_{\mu}e_{L}-\overline{\nu}\gamma_{\mu}\nu\right)
-(\sin{\theta})^{2}\overline{e}\gamma_{\mu}e\right]\right]\gamma^{\mu}\nu+\\
&+\frac{g}{\sqrt{2}}
\left[\frac{3\sqrt{2}\left[1-4(\sin{\theta})^{2}\right]}{32g\left(\sin{\theta}\right)^{2}}
\overline{e}_{L}\gamma_{\mu}\nu\right]\gamma^{\mu}e_{L}=0
\label{Dirac}
\end{eqnarray}
in which the parameter $g$ is introduced for generality and the definition of the parameter $q=g\sin{\theta}$ is arbitrary as well, and it is precisely because these parameters are free that we choose them in order to eventually write these interactions in such a way that the comparison with the weak interactions can be drawn visually; with this choice no artificial difference is manifest and whether the two types of interactions have the same structure will be evident: indeed it is now easy to see that these interactions among spinors have precisely the same form of the weak interactions among leptons, as we are going to show next.
\section{The Weak Forces for Leptons}
In what we have done up to now, we have been able to obtain a system of Dirac field equations in the form above which can be written as
\begin{eqnarray}
\nonumber
&i\gamma^{\mu}\nabla_{\mu}e-\frac{3}{8}(\cos{\theta})^{2}\overline{e}\gamma_{\mu}e\gamma^{\mu}e
+q\tan{\theta}Z_{\mu}\gamma^{\mu}e-\\
&-\frac{g}{2\cos{\theta}}Z_{\mu}\gamma^{\mu}e_{L}+\frac{g}{\sqrt{2}}W^{*}_{\mu}\gamma^{\mu}\nu=me\\
&i\gamma^{\mu}\nabla_{\mu}\nu+\frac{g}{2\cos{\theta}}Z_{\mu}\gamma^{\mu}\nu
+\frac{g}{\sqrt{2}}W_{\mu}\gamma^{\mu}e_{L}=0
\end{eqnarray}
upon definition of the following vectors
\begin{eqnarray}
&Z_{\mu}=\frac{3\cos{\theta}}{8g\left(\sin{\theta}\right)^{2}}
\left[\frac{1}{2}\left(\overline{e}_{L}\gamma_{\mu}e_{L}-\overline{\nu}\gamma_{\mu}\nu\right)
-(\sin{\theta})^{2}\overline{e}\gamma_{\mu}e\right]
\label{neutral}\\
&W_{\mu}=\frac{3\sqrt{2}\left[1-4(\sin{\theta})^{2}\right]}{32g\left(\sin{\theta}\right)^{2}}
\overline{e}_{L}\gamma_{\mu}\nu
\label{charged}
\end{eqnarray}
where the torsional interactions among spinors are formally identical to the weak forces for leptons and for which it is in terms of lepton bound states that the weak vector mediators are built, and there is no Higgs field whether composite or not appearing at all: notice, first, that as the interactions among spinors have precisely the form of the weak forces for leptons then in both approaches the phenomenology of the effective interactions for the couple electron-neutrino will yield the same results; further, we have that here it is in terms of bound states of leptons that the weak vector mediators emerge as dependent degrees of freedom whereas in the standard model the fundamental vectors mediating the weak interactions are independent degrees of freedom, and therefore here the weak forces are the effect of the presence of torsion while in the standard model the weak interactions are dynamical fields; finally, in the present model all fields already have the correct amount of degrees of freedom so that the Higgs boson is not needed whereas in the standard model the vector bosons must get additional degrees of freedom transferred to them from the Higgs boson, and therefore in the present model the Higgs boson is useless while in the standard model the Higgs boson is necessary as a reservoir of degrees of freedom.

According to the expressions (\ref{neutral}-\ref{charged}) we have the following
\begin{eqnarray}
&\nabla_{\mu}Z^{\mu}=-\frac{3\cot{\theta}}{16}\frac{m}{q}
\left(i\overline{e}\gamma e\right)
\label{conservedneutral}\\
&\nabla_{\mu}W^{\mu}=-\frac{3\sqrt{2}\left[1-4(\sin{\theta})^{2}\right]}{32\sin{\theta}}
\frac{m}{q}\left(i\overline{e}\gamma\nu\right)
\label{conservedcharged}
\end{eqnarray}
for which the correct sign of the partially conserved axial currents is obtained when the $\theta$ angle is smaller than $\frac{\pi}{6}$ radians as observations tell: remark that as there cannot be mixing between electron and neutrino then there cannot be mixing between these weak vector bosons compatibly with the fact that the relationships $Z^{2}\leqslant0$ and $W^{2}\leqslant0$ ensure they already have the three degrees of freedom they need to be massive; once again we stress that the main advantage of the present model is definitely that its asymmetric and massive, effective weak vector bosons are present with no Higgs field necessary.
\section*{Conclusion}
In this paper, we have been able to prove that it is possible to consider the Dirac field equations and see that the torsional interactions for spinors can be written in the form of the weak forces for leptons; this form is obtained by taking the electron and neutrino field to construct the weak vector mediators, and as a consequence of this fact the vector bosons do not have any gauge transformations already having the amount of degrees of freedom they need to be massive, and the Higgs field is therefore not needed: in this way a Higgsless standard model has been constructed, and in it the weak forces among leptons are effect of the presence of torsion among spinors. Thus the information about the weak forces is contained within torsion, in a similar way as the information about gravitation is contained within the metric and the information about electrodynamics is contained within the gauge potentials; this suggests that the weak forces, gravity and electrodynamics although intrinsically different fields may be accommodated into different parts of the same connection that defines the most general dynamics of spinors. This is different from what we usually call unification, and yet it clearly is a conceptual unification \cite{s-s,s-s/1,f/1}.

On the other hand, the most important issue left open by this model regards the smallness of the coupling, since these interactions are of torsional origin and the torsion is usually restrained by very stringent limits \cite{a-p,k-r-t,h-a-c-c-s-s}: however, the constraining bounds are placed in terms of vacuum models that do not apply to the context of the present discussion; nevertheless, we have that the interactions due to torsion are supposed to be manifest at the Planck scale, although for this approach to work they must be manifest already at the Fermi scale: that these effects may be relevant at these energies is only possible with an energy-dependent coupling, although we do not know if such an energy-dependent coupling is viable. An extensive further study about this possible running coupling is at the moment being carried off.

\end{document}